\documentclass[12pt,a4paper]{article}

\usepackage{times}
\usepackage{graphicx}
\usepackage{amssymb,amsmath}
\usepackage{mathrsfs}
\usepackage{fancybox}
\usepackage[latin1]{inputenc}
\usepackage{pst-all}
\usepackage{epsfig}
\usepackage{rotating}
\usepackage{subfig}
 
\newcommand{\be}{\begin{equation}}
\newcommand{\ee}{\end{equation}}
\newcommand{\bea}{\begin{eqnarray}}
\newcommand{\eea}{\end{eqnarray}}
\newcommand{\lb}{\label}
\newcommand{\bdm}{\begin{displaymath}}
\newcommand{\edm}{\end{displaymath}}

\newcommand{\I}{{\rm i}}

\begin{document}

\begin{titlepage}

\noindent
\begin{center}
\vspace*{1cm}

{\large\bf DOES THE QUANTUM MECHANICAL WAVE FUNCTION
  EXIST?}\footnote{Published in: {\em Zagadnienia Filozoficzne w
    Nauce}, {\bf 66}, 111--128 (2019).} 

\vskip 1cm

{\bf Claus Kiefer} 
\vskip 0.4cm
Institute for Theoretical Physics, University of Cologne
\vspace{1cm}

\begin{abstract}
I address the question whether the wave function in quantum theory
exists as a real (ontic) quantity or not. 
For this purpose, I discuss the essentials of the quantum formalism and emphasize the
central role of the superposition principle. I then explain the measurement
problem and discuss the process of decoherence. Finally, I address the
special features that the quantization of gravity brings into the
game. From all of this I conclude that the wave function really
exists, that is, it is a real (ontic) feature of Nature.  
\end{abstract}

\end{center}

\end{titlepage}


\section{Quantum theory}

The title of my contribution may sound somewhat surprising, at least
at first glance. After all, the quantum mechanical wave function and
its generalizations in quantum field theory (generically here called 
$\Psi$) are standard tools in
quantum theory and its many applications in physics, chemistry, and
even biology. This is true, and one can definitely say that $\Psi$
exists in a mathematical sense. The question addressed here instead refers to
whether $\Psi$ can be attributed an ontic or merely an epistemic role,
that is, whether $\Psi$ can be attributed reality in the same way as,
for example, an electric field possesses, or whether it merely describes
something like an information catalogue (as Schr\"odinger once put
it). This is a question that has occupied physicists since the advent
of quantum theory in the 1920s and that still occupies them today;
see, for example, d'Espagnat (1995), Kiefer (2015a), or Boge (2018) and the many
references quoted therein. Here, I will argue that that the answer to
the question posed in the title is definitely in the affirmative, and I
will try to put together the main arguments of why this is so and why
the wave function has an ontic (real) status. Some of these arguments
have been presented in an earlier article (Kiefer, 2012b), to which I
will occasionally refer.  

At the heart of all of quantum theory is the {\em superposition
  principle}. It can be separated into a kinematical and a dynamical
version (Joos et al., 2003). The kinematical version expresses the fact that if
$\Psi_1$ and $\Psi_2$ are physical states, then 
$\alpha\Psi_1+\beta\Psi_2$ is again a physical state, where $\alpha$
and $\beta$ are complex numbers.  
For more than one degree of freedom, this leads to the important
concept of
{\em entanglement} ({\em Verschr\"ankung}) between systems (Kiefer,
2015a), which plays a particular important role in modern developments
such as quantum information. The very concept of a quantum computer
relies on entanglement. 

It is clear that this kinematical version only makes sense if it is
consistent with the dynamics of the theory. But this is the case. The
fundamental equation is the Schr\"odinger equation (by which I include
its field theoretic generalization, the functional Schr\"odinger
equation), and this equation is {\em linear}: the sum of two
  solutions is again a solution. An importance consequence of the
  superposition principle is obvious: the space of what we may call 
``classical states'' form only a tiny subset in the space of all
possible states. A simple example is the superposition of two
localized states, each of which can describe a classical state, to a
nonlocal (and thus nonclassical) state. It must be emphasized that the
quantum mechanical wave functions are not defined on spacetime, but on
{\em configuration space} (cf. e.g. Zeh, 2016 for a lucid conceptual
discussion). Except for the case of one particle, this is a
high-dimensional space: the dimension is $3N$ for a system of $N$
particles, and infinite in field theory.
Otherwise, there would be no entanglement between systems.

Entanglement is the central distinguishing feature of quantum theory.
As already Erwin Schr\"odinger put it (Schr\"odinger, 1935, p.~555):

\begin{quote}
I would not call that {\em one} but rather {\em the}
characteristic trait of quantum mechanics, the one that enforces
its entire departure from classical lines of thought. By the
interaction the two representatives (or $\psi$-functions)
have become entangled. \ldots Another way of expressing the
peculiar situation is: the best possible knowledge of a
{\em whole} does not necessarily include the best possible
knowledge of all its {\em parts}, even though they may be
entirely separated \ldots
\end{quote}

The superposition principle has been experimentally tested
in uncountably many experiments (Schlosshauer, 2007; Kiefer, 2015a).
Even before the term entanglement was coined, it was clear
that the electrons in a helium atom must be
entangled in order to lead to the correct observed binding energies 
(Hylleraas, 1929). Modern experiments include the interference of
biomolecules, the entanglement of photon pairs over distances of
hundreds of kilometres, and the observation of neutrino oscillations,
to name only a few; see, for example, Deng {\em et al.} (2019)
for an experiment involving interference between light sources
separated by 150~million kilometres. There can thus be no doubt that
the superposition 
principle holds. The generation of ``macroscopic'' superpositions is
being seriously considered; see, for example, Clarke and Vanner (2018). 
 
In the mathematical language of quantum theory,
the validity of the superposition principle is encapsuled in the use
of vector spaces for the quantum states (wave functions). The stronger
concept of a Hilbert space (using a scalar product between states) is
motivated by the probability interpretation of quantum theory, which
by itself is connected to the ``measurement problem'' discussed since
the early days of the theory. This measurement problem refers, in
fact, to the only class of situations in which the superposition
principle seems to break down.

What is the measurement problem? Let us consider the simple situation
of an apparatus, A, coupled to a system, S:\footnote{This and the
  following diagramme are taken from our monograph (Joos et al., 2003).}
\begin{figure}[h]
\begin{center}
\setlength{\unitlength}{1cm}
\begin{picture}(5,1) \thicklines 
\put(0,0){\framebox(2,1){S}}
\put(3,0){\framebox(2,1){A}}
\put(2,0.5){\vector(1,0){1}}
\end{picture}
\end{center}
\end{figure}
I emphasize that both system and apparatus are described by quantum
theory. This analysis goes back to John von Neumann (von Neumann,
1932). The simplest situation of an interaction is the ``ideal
measurement'': the system is not disturbed by the apparatus, but the
state of the apparatus becomes correlated with the state of the
system. If S is in a state $|n\rangle$ and A in an initially
uncorrelated state $\Phi_0\rangle$, the total state of S and A evolves as
\be
|n\rangle|\Phi_0\rangle \stackrel{t}{\longrightarrow}
|n\rangle|\Phi_n(t)\rangle.
\ee
The measurement problem appears when we consider a superposition of
possible states $|n\rangle$.\footnote{In the simple situation of
  spin-1/2, one would have two states  $|n\rangle$, one corresponding
  to (say) spin up and the other to spin down.}
This leads to the evolution
\be
\lb{super}
\left(\sum_n c_n|n\rangle\right)|\Phi_0\rangle
    \stackrel{t}\longrightarrow\sum_n c_n|n\rangle
    |\Phi_n(t)\rangle,
\ee
where $|\Phi_n(t)\rangle$ is the resulting state (`pointer state') of
A. 
But \eqref{super} is a {\em macroscopic} superposition! 
Since such superpositions\footnote{An especially impressive example is
  Schr\"odinger's cat.} are not observed, John von Neumann postulated
the occurrence of a ``collapse of the wave function'' in
measurement-like interactions; he did not, however, present a
dynamical equation for such a collapse, which must be unitarity violating
and is thus in conflict with the Schr\"odinger equation.

In more recent years, various models of wave function collapse have
been presented in the literature and possible experimental tests have
been discussed.\footnote{See, for example, Bassi et al. (2013) for a
  comprehensive review.} It must be emphasized that most of these models only
make sense if the wave function acquires a real (ontic) status. This
is different from its role in the Copenhagen interpretation of quantum
theory, where the `collapse' has the mere formal meaning of an
information increase. We shall see in the next section how we can
proceed without assuming a dynamical collapse, that is, without
violating the unitarity of quantum theory. 

Before doing so, I want to conclude this section with some remarks on
relativistic quantum theory, in particular the Dirac equation. 
The Dirac spinor appearing there should not be
confused with the wave function discussed above. The spinor is {\em
  not} defined in configuration space; it is defined on a classical
(in general four-dimensional) spacetime. It thus cannot describe
entanglement and can only serve to address one-particle situations;
it can describe correctly the situation in the hydrogen atom, but
cannot even be formulated for the helium atom.\footnote{Cf. in this
  context Zeh (2016).} Relativistic quantum theory is only consistent
in the form of quantum {\em field} theory; the Dirac equation follows
from quantum electrodynamics (QED) for the special case of
one-particle excitations.      
When we talk here about the ontological status of $\Psi$, this refers in
the general case of quantum field theory to wave {\em functionals}.
These functionals are defined on the configuration space of all
fields; in the case of QED, for example, this is the space of all
vector potentials and charged Grassmann (anti-commuting) fields.

\section{Decoherence}

How can one understand the nonobservation of superpositions such as
\eqref{super} without advocating an explicit collapse?
The key role in answering this question is played by the presence of
the ubiquitous environment for the apparatus. This was clearly
recognized in the pioneering work by H.-Dieter Zeh in
1970.\footnote{The original reference is Zeh (1970). See Joos et
  al. (2003) for details and references.} 
`Environment' is here a technical terms that stands for additional
degrees of freedom whose interaction with the `apparatus' (or other
systems under consideration) cannot be avoided. In concrete examples, 
these may be air molecules or photons that scatter off the
`apparatus'. One thus has instead of the above diagramme the 
following situation:

\begin{figure}[h]
\begin{center}
\setlength{\unitlength}{1cm}
\begin{picture}(5,1)(-0.25,-0.25) \thicklines 
\put(0,0){\framebox(1,0.5){S}}
\put(1.5,0){\framebox(1,0.5){A}}
\put(1,0.25){\vector(1,0){0.5}}
\put(3,-0.25){\framebox(1.5,1){E}}
\put(2.5,0.15){\vector(1,0){0.5}}
\put(2.5,0.25){\vector(1,0){0.5}}
\put(2.5,0.35){\vector(1,0){0.5}}
\put(-0.25,-0.25){\dashbox{0.1}(3,1){}}
\end{picture}
\end{center}
\end{figure}

Here, E stands for the environmental degrees of freedom, and the three
arrows between A and E indicate the many degrees of freedom.

If the environment is in an initial state $|E_0 \rangle$,
 the superposition principle for the whole system of S, A, and E leads to
\be
\lb{super2}
\left( \sum_n c_n|n\rangle|\Phi_n\rangle\right)|E_0 \rangle
  \quad  \stackrel{t}\longrightarrow \quad
   \sum_n c_n|n\rangle |\Phi_n\rangle |E_n\rangle. 
\ee
But this is an even more macroscopic superposition than \eqref{super}
because it not only includes system and apparatus, but also the many
degrees of freedom of the environment. Has the situation not become
worse now? The answer is no. The reason is because the degrees of
freedom of the environment are in general not accessible; when
dealing with S and A only, one has to trace them out and to consider
instead the reduced density matrix of S and A alone, from which all
{\em local} observations follow. Since different environmental states
are in general orthogonal (because they can discriminate between
different states of A), $\langle E_n|E_m \rangle \approx \delta_{nm}$,
the reduced density matrix assumes the form 
\be
\lb{rho}
\rho_{\rm SA} \approx \sum_n |c_n|^2 |n\rangle\langle n|
   \otimes |\Phi_n \rangle \langle \Phi_n |,
 \ee
which is approximately (but not identically) equal to a classical
stochastic mixture. The information about the original superposition
of \eqref{super} has now been transferred to a nonlocal correlation
between S and A on the one side, and E on the other side. They are no
longer observable at S and A itself:
``The interferences exist, but they are not {\em there}.''
The various system states $|n\rangle$ are distributed with
probabilities $|c_n|^2$ according to the Born rule of quantum theory. 
(It should be noted, though, that the very notion of density matrix is
based on the validity of the Born rule.) 

This irreversible emergence of classical properties (nonobservability
of interference terms) through the unavoidable interaction with the
environment is called {\em decoherence}. It has been explored in the
last decades, both experimentally and theoretically.\footnote{Major
  reviews are Joos et al (2003), Zurek (2003), Schlosshauer
  (2007). Crucial experiments are also discussed in Haroche (2014).}  
According to decoherence, macroscopic objects {\em appear}
classically, although they are fundamentally described by quantum
theory. Decoherence is a process that can be treated entirely within
standard quantum mechanics and which can be based on realistic
processes discusses in a quantitative manner.\footnote{Such
  quantitative calculations were presented for the first time in Joos
  and Zeh (1985).} 

What are the consequences of this for the interpretation of quantum
theory in general and for the wave function in particular?
If the superposition principle and the Schr\"odinger equation are
universally valid, one arrives at what is called the Everett or
many-worlds interpretation (see e.g. d'Espagnat, 1995; Zeh, 2016). Unitary
quantum theory is then exact and never violated. The dead and the
alive Schr\"odinger cat, for example, then indeed exist simultaneously
in different ``Everett branches'', and also the observer seeing the
cat exists in two versions. In this point of view, the wave function
definitely has an ontic status and exists in the way discussed above.
The Everett interpretation together with decoherence makes the
measurement problem obsolete.

A question often asked is about the derivation of the probability
interpretation (the Born rule) in the Everett picture. This has been
discussed at length in the literature; see, for example, Zurek (2018)
and the references therein. The probability interpretation only makes
sense for situations in which decoherence is effective, because only
then the various alternatives can be treated independently and can be
assigned probabilities. Whether the Born rule can then really be {\em
  derived} or only be made plausible, is a contentious issue. But what is
clear that the Everett interpretation together with decoherence and
the Born rule gives a consistent picture that is not in need
of completion. 

The Everett interpretation is the simplest one at the level of the
mathematical formalism. The fundamental equations are all linear. It
is not a simple interpretation if one sticks to a classical picture of
the world. This is what the main alternative -- explicit collapse
models -- wants to achieve (see e.g. Bassi et al, 2013). But this
requires a modification of the usual formalism by bringing in
nonlinearities or stochastic terms. Also here, the wave function
assumes an ontic status. The main task is to work out concrete models
and to explore their experimental status. 

A rather mild modification is the de~Broglie--Bohm
theory. The Schr\"odinger for $\Psi$ is left untouched, but
in addition particle (or classical field) configurations are
introduced. The wave function, which is defined in configuration space,
acts as a kind of `guiding field' for the particles in ordinary
space. There, too, it has an ontic status and can thus be assumed to
exist. At least in nonrelativistic quantum mechanics, the predictions
of the  de~Broglie--Bohm theory agree with the predictions of standard quantum
theory.

The prototype of an epistemic point of view is the Copenhagen
interpretation. There, $\Psi$ merely provides an increase of information
during a measurement and has no physical existence on its own -- only
the classical concepts such as particle posititions have. But is such
a point of view really consistent and satisfactory?  This is hard to
believe. New light on these interpretational questions is shed by
entering the realm of quantum gravity and quantum cosmology. This is
the topic of my final section.

\section{Quantum gravity}

In 1957, a group of distinguished physicists met at the University of
North Carolina to explore the prospects of gravitational physics.
This also included the possible quantization of the
gravitational field. In the discussion, Richard Feynman came up with
the following 
gedanken experiment. In a Stern--Gerlach type of setting, a particle
is brought into a superposition of, say, spin up and spin
down. Introducing some interconnections to a macroscopic object, say
a ball of 1~cm diameter, one can bring the ball into a superposition of being
translated upwards and downwards. But this corresponds to a
superposition of two measurable gravitational fields (measurable e.g. with a
Cavendish balance). Feynman then states (DeWitt, 1957):

\begin{quote}
\ldots if you believe in quantum mechanics up to any level then you
have to believe in gravitational quantization in order to describe
this experiment. \ldots It may turn out, since we've never done an
experiment at this level, that it's not possible \ldots that there is
something the matter with our quantum mechanics when we have too much
{\em action} in the system, or too much mass---or something. But that
is the only way I can see which would keep you from the necessity of
quantizing the gravitational field. It's a way that I don't want to
propose. \ldots 
\end{quote} 

\begin{figure}[h]
   \includegraphics[width=0.8\textwidth]{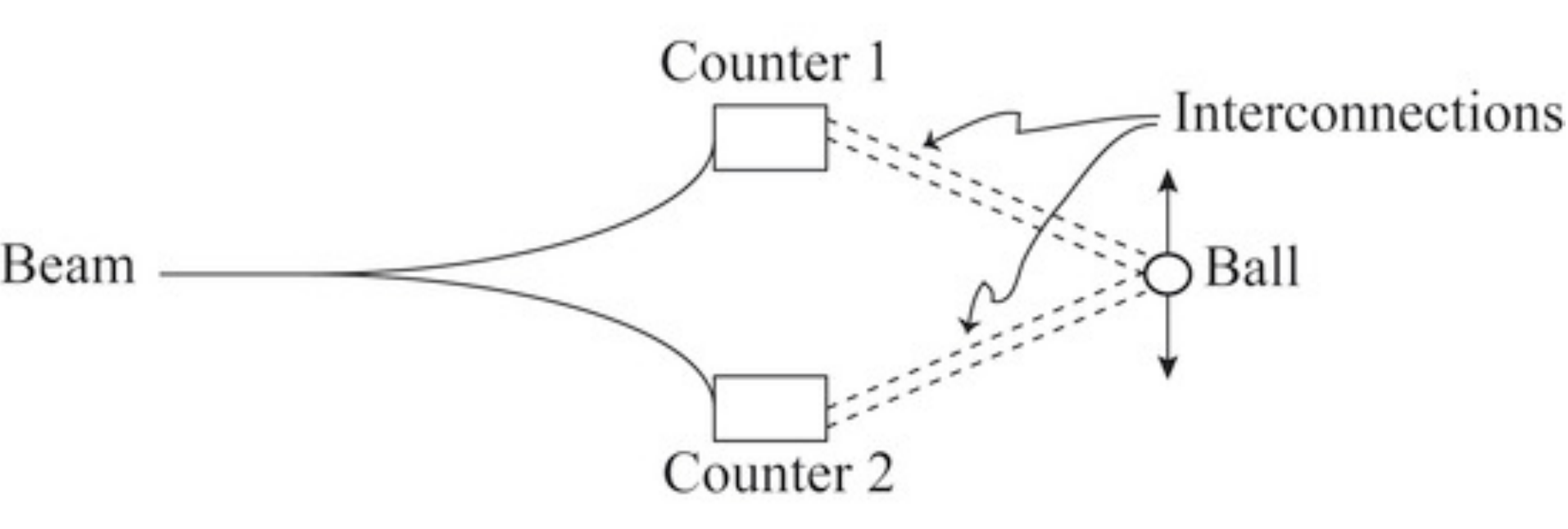} 
\caption{Feynman's gedanken experiment in which a microscopic
  superposition is transferred to a macroscopic one of a ball being
  simultaneously at two different places. Figure adapted from DeWitt, 1957.}
\end{figure}

In order words, unless one assumes that the superposition principle
and the standard formalism of quantum theory is violated when
gravitational fields play a role (as, for example, Lajos Di\'osi and
Roger Penrose 
envisage), the quantization of gravity seems unavoidable. The majority
of researchers thus accepts the assumption of 
extrapolating the standard linear formalism of quantum theory to
quantum gravity. This holds for almost all of the existing approaches,
from canonical and covariant quantum gravity up to string theory
(Kiefer, 2012a).

At present, there is a discussion about the possibility to observe
gravitational superpositions in the laboratory. There are proposals
to probe a nonclassical gravitational field generated by
two masses each of which is superposed at two locations (see e.g. Marletto and
Vedral~2017) or to probe such a field generated by the superposition
of one mass in the spirit of Feynman's proposal cited above
(see e.g. the remarks in Schm\"ole {\em et al.}~2016). The
observability of such superpositions also meets with criticism 
(Anastopoulos and Hu~2018).

What are the consequences of quantum gravity for our question about the
reality of the wave function? In order to answer this question, it is
sufficient to use the simplest and most conservative approaches to
quantum gravity, which is quantum geometrodynamics.\footnote{Details
  and relevant references can be found, for example, in my monograph
  (Kiefer, 2012a).} One arrives at this theory when asking the
following question: what is the quantum formalism that gives back
Einstein's equations in the semiclassical (WKB) limit? This is analogous
to the heuristic procedure that Erwin Schr\"odinger led to his
equation in 1926.

The canonical formalism of general relativity discloses the real
dynamical quantity of the theory: it is the {\em three-dimensional}
geometry. The configuration space is thus the infinite-dimensional
space of all three-dimensional metrics, with an additional constraint
which guarantees that metrics related by coordinate transformations are
counted only once. The theory possesses four local constraints, which
after Dirac quantization are heuristically transformed into quantum
constraints on physically allowed wave functionals. 
In a shorthand notation, they read
\be
\label {wdw}
\hat{H}\Psi=0,
\ee
where $\hat{H}$ denotes the Hamilton operator of all gravitational and
nongravitational degrees of freedom. The functional $\Psi$ is defined
on the space of three-metrics and nongravitational fields. Equation
\eqref{wdw} is also called Wheeler--DeWitt equation.\footnote{More
  precisely, if written out, \eqref{wdw} includes the Wheeler--DeWitt
  equation and the diffeomorphism constraints.} 

One recognizes from \eqref{wdw} the absence of any external time parameter 
(see in this context Kiefer, 2015b). This is obvious for conceptual
reasons. In classical relativity, spacetime (four-geometry) plays the
same role that a particle trajectory plays in mechanics. After
quantization, spacetime has disappeared in the same way as the
particle trajectory has disappeared in quantum mechanics. But whereas
in quantum mechanics Newton's absolute time $t$ has survived, 
no such absolute time is present in Einstein's theory.  
As a result, the fundamental quantum gravity equations are timeless.

Of special concern here is quantum cosmology -- the application of
quantum theory to the Universe as a whole. In the simple case of a
Friedmann universe with scale factor $a$ and a conformally coupled
scalar field $\chi$, the Wheeler--DeWitt equation assumes the form
(after some rescaling and with a suitable choice of units):
\be
\lb{qc}
\hat{H}_0\psi(a,\chi)\equiv (-H_a+H_{\chi})\psi\equiv
\left(\frac{\partial^2}{\partial a^2}-\frac{\partial^2}{\partial\chi^2}
-a^2+\chi^2\right)\psi=0.
\ee

How can one interpret such equations? At the most fundamental level, 
there is no time and there are no classical observers who could
perform measurements. Therefore, the Copenhagen interpretation which
requires the need of classical measurement agencies from the outset,
is inapplicable.  
The question thus arises in which limit
approximate notions of time and observers (more generally, of classical
properties) emerge and what relevance this emergence has for the
interpretation of the wave function. 

Such a limit exists and is well understood (Kiefer, 2012a, 2015b). It
is similar to the Born--Oppenheimer approximation in molecular
physics. If one adds inhomogeneous degrees of freedom to the
Hamiltonian in \eqref{qc}, the Wheeler--DeWitt equation is of the form
\be 
\left(H_0+\sum_nH_n(a,\phi,x_n)\right)\Psi(\alpha,
 \phi,\{ x_n\})=0,
\ee
where the $x_n$ stand for the inhomogeneities (gravitational waves,
density perturbations). 
Writing $\Psi=\psi_0\prod_n\psi_n$ and assuming that
$\psi_0$ is of WKB form, that is, $\psi_0\approx C\exp(\I S_0/\hbar)$
(with a slowly varying prefactor $C$), one gets
\be
\I\hbar\frac{\partial\psi_n}{\partial t} \approx H_n\psi_n
\ee
with
\bdm
\frac{\partial}{\partial t}:= \nabla S_0\cdot\nabla.
\edm
This is nothing but a set of time-dependent Schr\"odinger equations
for the inhomogeneities with respect to a time variable $t$ that is
defined from the homogeneous cosmological background;
$t$ is also called `WKB time' and controls the dynamics in this
approximation. Only in this limit can one talk about the probability
interpretation of quantum theory and the existence of observers. It is
thus not at all obvious whether the standard notion of Hilbert space
need, or even can, be extrapolated to the level of full quantum
gravity (beyond this level of the Born--Oppenheimer approximation).

In quantum cosmology, 
 arbitrary superpositions of the gravitational
field and matter states can occur. How can we understand the emergence
of an (approximate) classical Universe? This is achieved by the
process of decoherence introduced above (Kiefer, 2012b). 
Decoherence is a process in
configuration space, and the irrelevant degrees of freedom can be
taken to be part of the inhomogeneities $x_n$. In this way, the scale
factor $a$ and the field $\chi$ can be shown to assume classical
properties. The same then holds for WKB time $t$, which is constructed
from these background variables. 
After this classicality is understood, one can investigate
decoherence for some relevant inhomogeneous degrees of freedom; these
include the inhomogeneities of an inflaton scalar field and
of the metric, giving rise, after decoherence, to the observed CMB
anisotropies and the (not yet discovered) primordial gravitational
waves. In all these considerations, the wave function is assumed to be
real (ontic); this is also the case if one applies collapse models to
quantum cosmology. I should also mention that even the problem of the
arrow of time can, at least in principle, be understood in the
framework of timeless quantum cosmology (Zeh, 2007).

It is clear that the debate about the correct interpretation of
quantum theory will continue, at least until a clear experimental
decision is reached (which could take quite a while). In this
contribution, I have collected arguments which strongly support the
point of view that the wave function is real (ontic), in the same way
as, say, an electric field, is real. Thus, the wave function {\em
  exists}. The perhaps most important open question is: what is the
configuration state for the wave functional at the most fundamental
level? In canonical quantum gravity, it is the space of
three-geometries plus nongravitational fields; what it is at the level
of a fundamental quantum theory of all interactions, is unknown.

{\em Note added for arXiv version}: At the conference, I did not talk
about the possibility to directly measuring the wave function. That
this is indeed possible, in a certain sense,\footnote{See
  J.~S.~Lundeen {\em et al.}, Direct measurement of the quantum
  wavefunction, {\em Nature}, {\bf 474}, 188--191 (2011). In this, the
notion of weak measurement plays a crucial role.}
supports my point that
the wave function has an ontic meaning. 


\section*{Acknowledgements}

I am grateful to the organizers of the conference
{\em On What Exists in Physics} (Krak\'ow, October~2017) for inviting
me to such an  
inspiring event. I am, in particular, indebted to Professor Micha\l\
Heller for many interesting discussions.



\end{document}